**Small gold nanorods with tunable absorption for photothermal microscopy in cells**

*Edakkattuparambil Sidharth Shibu[1,2], Nadezda Varkentina[1,2], Laurent Cognet[1,2] and Brahim Lounis[1,2],\**

((Optional Dedication))

Dr. Edakkattuparambil Sidharth Shibu, Dr. Nadezda Varkentina, Dr. Laurent Cognet and Prof. Brahim Lounis
[1]University of Bordeaux, LP2N, F-33405 Talence, France
[2]Institut d'Optique & CNRS, LP2N, F-33405 Talence, France
Email: brahim.lounis@u-bordeaux.fr

Over the last decade, single-molecule optical microscopy has become the gold-standard approach to decipher complex molecular processes in cellular environments.[1-3] Single-molecule fluorescence microscopy has several advantages such as ease of application, high sensitivity, low invasiveness and versatility due the large number of available fluorescent probes. It bears however some drawbacks related to the poor photostability of organic dye molecules[4] and auto-fluorescent proteins[5-7] and and to the relatively large size of semiconductor nanoparticles in the context of live cell applications.[4,8,9] The overall size of the functional biomarkers is a general issue for any imaging approach because of steric hindrance effects in confined cell regions. Small red-shifted nano-emitters that are highly photostable are not currently available, while they would combine the best physical and optical penetration properties in biological tissues. Although single-molecule absorption microscopy was early used to detect single-molecules[10] at cryogenic temperatures, it is only with the advent of photothermal microscopy[11,12] that practical applications of absorption microscopy were developed in single-molecule research. Photothermal imaging (PhI) microscopy can indeed reach unprecedented sensitivities for detecting tiny absorbers with absorption cross-section as small as a few $10^{-16}$ cm$^2$.[12-14]

Several types of nano-objects ranging from gold nanoparticles,[15-19] carbon nanotubes[20,21] to quantum dots[22] are detected at the single-particle level leading to refined spectroscopic studies and ultra sensitive imaging applications. PhI displays extremely stable signals and is therefore appealing for biological applications. Furthermore photothermal microscopies are totally insensitive to non-absorbing scatterers, even when large objects with strong refractive index contrasts are present within the surroundings of the imaged nanoparticles, as is often the case in biological samples.[23] Several live cell application of PhI involve the detection of small gold nanospheres in order to perform long term single molecule tracking.[9,24,25] A limitation however arises due to background signal from endogenous cellular components, since these nanoparticles have to be excited at their plasmon resonance, which lies around 530 nm. Although one can also take advantage of this intrinsic absorption to achieve label-free imaging,[26] very small red-shifted absorbing nanoparticles would represent better probes for PhI since near-infrared is a region where the absorption of cell organelles is negligible. The use of gold nanorods as small probes absorbing in the near infra-red, is a promising strategy as they would combine good subcellular accessibility,[9] low contribution from intrinsic cellular signals and perfect photostability.

Gold nanorods are elongated nanoparticles, which display additional red-shifted plasmon resonance as compared to their spherical counterparts.[27] More precisely, transverse or longitudinal plasmon resonances result from the electron cloud oscillations excited by light polarized along their short-axis (diameter) or long-axis (length) respectively. Interestingly, the short axis plasmon resonance is basically independent on nanorods dimensions, while the long-axis plasmon resonance can be tuned over a wide spectral range (from 620 nm to 1100 nm) based on their aspect ratio. Although large diameter nanorods (> 8 nm) have been extensively used in different biological applications, such as bioimaging, delivery, sensing and therapy,[18,28,29] the synthesis of nanorods with smaller diameter is more delicate and are more recent.[30] Here, we report the synthesis, sorting and characterization of mono-disperse gold nanorods. These nanorods display tunable red-shifted plasmon resonance as compared to their spherical counterparts. A dual color PhI microscope was developed to image them down to the single nanorod level and to demonstrate that nanorods are promising basic building blocks for the realization of cellular imaging biomarkers in the near infrared.

Nanorods were synthesized using a modified pH-prompted seedless protocol (see experimental section).[30] The as-prepared solution was first characterized using UV/Vis absorption spectroscopy from 400 nm to 1000 nm (**Figure 1A**). The absorption spectrum shows the characteristic splitting of the surface Plasmon resonance (SPR) into two modes in case of gold nanorods (transverse at 516 nm and longitudinal ~ 680 nm).[31-33] Nanorod morphologies were then characterized by transmission electron microscopy (TEM). All obtained nanorods had lengths smaller than 50 nm and diameters smaller than 7 nm (**Figure 1B**). More precisely, the distributions of nanorod lengths, diameters and aspect ratios were constructed (**Figure 1C-E**) revealing that the as-prepared solution was poly-disperse in length, and rather mono-disperse in diameter (6± 1nm).

In order to improve nanorods mono-dispersity with tunable longitudinal plasmon resonance, we performed length sorting by density gradient ultracentrifugation[34,35] (see experimental section). Upon centrifugation of the initial solution, distinct colored bands were visible (**Figure 2A**). Different fractions were collected and analyzed. **Figure 2B** displays UV/Vis absorption spectra of the 6 first collected fractions corresponding to the shortest nanorods. A manifold of narrower longitudinal SPR peaks is found as compared to the parent solution while the transverse SPR peaks are almost identical. This indicates that efficient sorting based on nanorod lengths was obtained which results in monodisperse nanorods solutions with tunable plasmon resonance peaks. In the following, we concentrated on the solutions containing the smallest nanorods. **Figure 2C** and **D** show TEM images of fractions with peak SPR at 618 nm and 642 nm respectively (image of a fraction containing longer nanorods is shown in **Figure S1**). From such images we constructed the histograms of nanorod lengths and aspect ratios (**Figure 2E-H**). We obtained average lengths of 10±1 nm (**Figure 2E**) and 13±1 (**Figure 2G**) and aspect ratios of 1.8±0.7 and 2.2±0.7 (**Figure 2F and 2H**) respectively. The latest are consistent with position of the SPR peaks measured in **Figure 2B**.

We next characterized by PhI microscopy the nanorod sorted samples at the single-particle level. PhI microscopy is indeed a sensitive imaging modality, which enables detection of nanometer-sized objects solely based on their absorption. It involves the detection of refractive index variations that are induced by photothermal effect in the local environment of an absorbing nanoparticle.[12,36-38] Gold nanoparticles of a few nanometers are efficiently detected by this imaging modality, thanks to their fast relaxation times (ps range) and large absorption sections around their plasmon resonance. In order to excite nanorods either at their transverse or at their longitudinal plasmon resonances, a PhI microscope was built with two-color excitation beams and a probe beam in the near infrared. The setup uses a low amplitude noise single frequency laser diode at 785 nm acting as probe beam (Innovative Photonic Solutions, TO-56) and a 640 nm laser (Coherent OBIS-FP) to excite nanorods at their longitudinal resonance or a 532 nm laser (Coherent Sapphire) to excite the nanorods at their transverse resonance (**Figure 3**). The intensities of the absorption beams were modulated at 450 kHz. The three beams were overlaid and focused onto the sample using a high NA objective (60x, NA=1.49). Absorption beams were circularly polarized to ensure that nanorods are equally excited regardless of their orientation in the sample plane. Upon light absorption by a nanorod, temperature elevation induces time-modulated variations of the refraction index in its close environment. The interaction of the probe beam with this index profile produces a scattered field with sidebands at the modulation frequency. The scattered field was then detected through its beat note with the probe field transmitted through the sample, which plays the role of a local oscillator and is extracted by lock-in detection. Samples were mounted on a piezo-scanner stage that allowed scanning the sample to acquire 2D images of nanorods. All data where acquired with integration times of 5 ms/pixel using a resolution of 100 nm/pixel. The setup also included white light trans-illumination and a CCD camera for bright-field imaging of the biological samples.

Samples were first prepared by spin coating a mixture of (1:1 ratio) CTAB stabilized nanorods and polyvinyl alcohol (PVA, 1.5%) on the surface of plasma cleaned glass slides. A drop of silicon oil was then added on the sample to ensure homogenous heat diffusion. PhI images of nanorod fraction with a SPR peak at 642 nm were acquired using laser excitation at 532 nm (**Figure 3B**) and 640 nm (**Figure 3C**) with the same intensities at the imaging plane. One to one correlation is found in the two images indicating that identical objects are detected at the two excitations wavelengths. The signal intensities originating from ~100 single nanorods excited at the two wavelengths were compared as shown in the histograms of **Figure 3D-E**. Several observations indicate that that single nanorods are detected at the two wavelengths. First, both distributions are narrow which also indicates that the dispersions of nanorod dimensions (both diameter and lengths) are very small within such length-sorted fractions. Second, signal dispersion at 640nm excitation also includes a tail towards the small signals attributed to a contribution from the out-of polarization-plane orientation of the rods. Indeed, for individual nanorods, absorption at 532 nm under circularly polarized excitation is weakly dependent on nanorod orientations[39] since the transverse mode is always excited (**Figure 3D**) while absorption at 640 nm circularly polarized excitation is dependent on nanorod orientation since the longitudinal mode of nanorods with long axis normal to the polarization plane are not excited. To further support that single nanorods are detected in **Figure 3B-C**, we also collected images with at 640 nm excitation beam having four different linear polarization orientations. As expected from single rod detection PhI signals strongly depend on the beam polarization **Figure 3G-I**. Noteworthy, nanorods oriented perpendicularly to the polarization axis display no PhI signal (see examples shown by arrowheads on **Figure 3G-I.**

We also note that a three fold higher intensity is observed under 640 nm excitation compared to 532 nm excitation at the peak of the histograms. This was expected for nanorods where the longitudinal SPR is more intense than the transverse one. Noteworthy in ensemble spectra presented in **Figure 2B**, the longitudinal SPR absorption peak is less than three times the transverse one due to orientation averaging of nanorods in solutions. PhI images and corresponding intensity histograms of other fractions are given in the supporting information (**Figure S2**). All of these fractions show similar trend in their intensity histograms, confirming that the synthesis protocol and DGU sorting leads to narrow size dispersion of nanorod solutions.

For bioimaging applications requiring very small nano-labels to access restricted cellular areas or complex tissue organizations, the signal to cellular background ratio at which small nanolabels can be detected is commonly the limiting factor. The smaller the nano-labels, the more difficult is the task due to the increasing weakness of detectable signal originating from the labels. In the case of PhI, intrinsic signal due to residual absorption of green light by cell mitochondria, lead to background signals that can reach that PhI signals of 5 nm gold nanoparticles excited at their plasmon resonance, i.e. around 532 nm. nanorods which we can detect with high signal to noise ratio at 640 nm while maintaining a small size, should thus represent a promising strategy in the context of cellular imaging with photothermal microscopy. In the following, we imaged nanorods in cellular environments. Fixed COS-7 cells were chosen because these cells were previously shown to display intense intrinsic PhI signals originating from mitochondrial light absorption. After fixation COS-7 cells were incubated with 5 nM solution of nanorods in PBS containing 3% bovine serum albumin followed by extensive rinsing steps to remove non-immobilized nanorods. **Figure 4** displays white light images of the cells and the corresponding PhI images recorded using 532 nm and 640 nm excitation. Accordingly to **Figure 3**, punctual signals originating from individual immobile nanorods can be identified on the cells in the two images. Importantly cellular (mitochondrial) structures are clearly visible under 532 nm excitation (**Figure S3**), which complicate the identification of nanorods around the mitochondria (arrow) at this excitation wavelength. In contrast, background signals originating from mitochondria are notably reduced under 640 nm excitation (**Figure S4**). In addition, individual nanorods display notably higher PhI signals 640 nm excitation as compared to 532 nm excitation, facilitating their detection of in cellular environments. The combination of these two effects thus participated to a clear enhancement of signal to cellular background ratio of detection of nanorods in cells.

In conclusion, we developed a novel strategy for photothermal imaging based on gold nanorods, which present strong optical absorption tunable from the red to the near IR. For biological applications, the use of these nanorods minimizes background signal from the cell organelles. We anticipate that they will constitute the next generation photothermal probe to study complex molecular dynamics in biological systems owing to their small size, tunable NIR-absorption, absolute photostability and chemical suitability for surface functionalization and bioconjugation.

**Experimental Section**

*Nanorod synthesis:* Briefly, $HAuCl_4.3H_2O$ (90 mL, 1 mM) was added to 270 mL cetyltrimethyl ammonium bromide (CTAB; 0.2 M) followed by $AgNO_3$ (9 mL, 4 mM). The solution was gently shaken. HCl (405 μL, 37%) was then added to adjust the pH to ~1. During this step, the color of the solution evolves from dark yellow to orange. Subsequently, 1260 μL of L-ascorbic acid (AA; 78.8 mM) was injected and gently shaken until colorless. Immediately, ice-cold $NaBH_4$ (540 μL, 10 mM) was slowly added and the solution was kept overnight at constant temperature (~30 °C). Appearance of dark blue color indicates the formation of nanorods. Excess CTAB was removed by centrifuging the parent solution at 20k rpm for 1.5h. The pellet was collected and re-dispersed in 10 mM CTAB (4 mL) solution.

*TEM imaging:* Monolayers of positively charged nanorods (due to CTAB encapsulating layer) were deposited and dried under ambient conditions on carbon coated Cu grids (200 mesh) that were negatively charged using a glow discharge technique (K950X Turbo evaporator with a 350X glow discharge head; Emitech, France).

*Nanorods length sorting by density gradient ultrahigh-centrifugation (DGU):* The density-gradient consisted of 4 layers of decreasing concentration of ethylene glycol (EG) from bottom to top (volume ratios, 80%, 70%, 60%, 50%) in 10mM CTAB, cooled at 4 °C before use. A 4 mL volume of nanorod solution was drop-casted on the top of cooled density gradient tube and centrifuged at a speed of 10k rpm for 3.5 h at 15 °C (Beckman Coulter ultrahigh centrifuge).

*Cell culture and fixation:* COS7 cells were cultured plated on #1 glass slides up to 60% confluence in DMEM medium supplemented with streptomycin (100 µg/ml), penicillin (100 U/ml), and 10 % bovine serum in a humidified atmosphere (95 %) at 5 % $CO_2$ and 37 °C. Cells were used for 12-14 passages and were transferred every 4 days. Before imaging, the cells were fixed in methanol at -30 °C, washed with phosphate buffered saline (PBS) and stored in PBS at 4°C until incubated with nanorods.

**Supporting Information**
Supporting Information is available from the Wiley Online Library or from the author.

**Acknowledgements**
This work was supported by CNRS, Agence Nationale de la Recherche (ANR-14-CE09-0018-01), Conseil Régional d'Aquitaine, the France-BioImaging national infrastructure (ANR-10-INBS-04-01) and IdEx Bordeaux (ANR-10-IDEX-03-02, LAPHIA) and SIRIC BRIO. E.S.S. acknowledges Marie Curie Individual Fellowship (655890 funding).

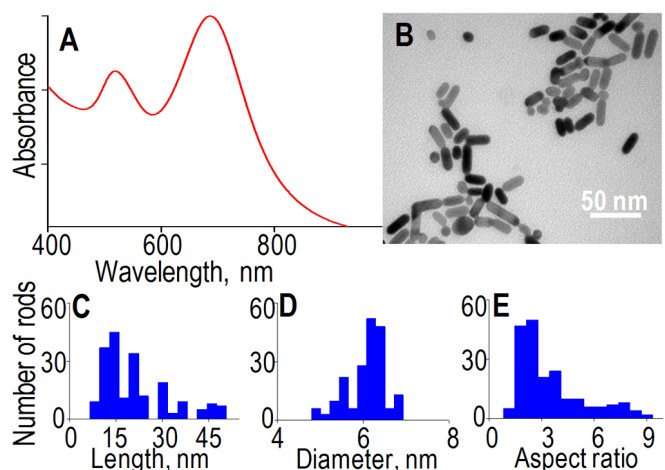

**Figure 1.** UV/Vis absorption spectrum (A), TEM image (B) and corresponding length (C), diameter (D) and aspect ratio (E) histograms of as prepared nanorods. Here, ~200 nanorods are used to construct the histograms.

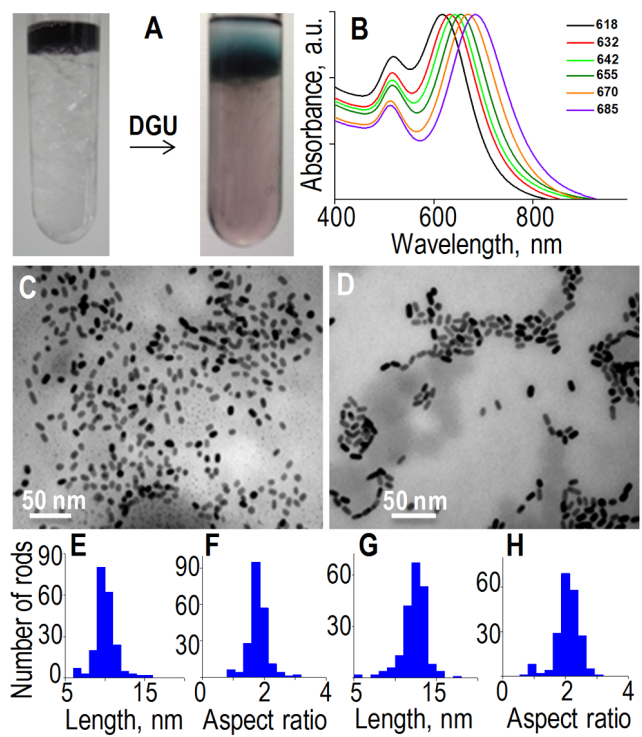

**Figure 2.** (A) Photographs of DGU tube before and after centrifugation. (B) UV/Vis absorption spectra of six fractions collected after DGU. (C-D) TEM images of two fractions (618 nm and 642 nm peak absorption). The corresponding length and aspect ratio histograms are presented in (E-H). They show average lengths of 10±1 nm (E) and 13±1 nm (G) with average aspect ratios of 1.8 (F) and 2.2 (H), respectively. Here, ~200 nanorods are used to construct histogram.

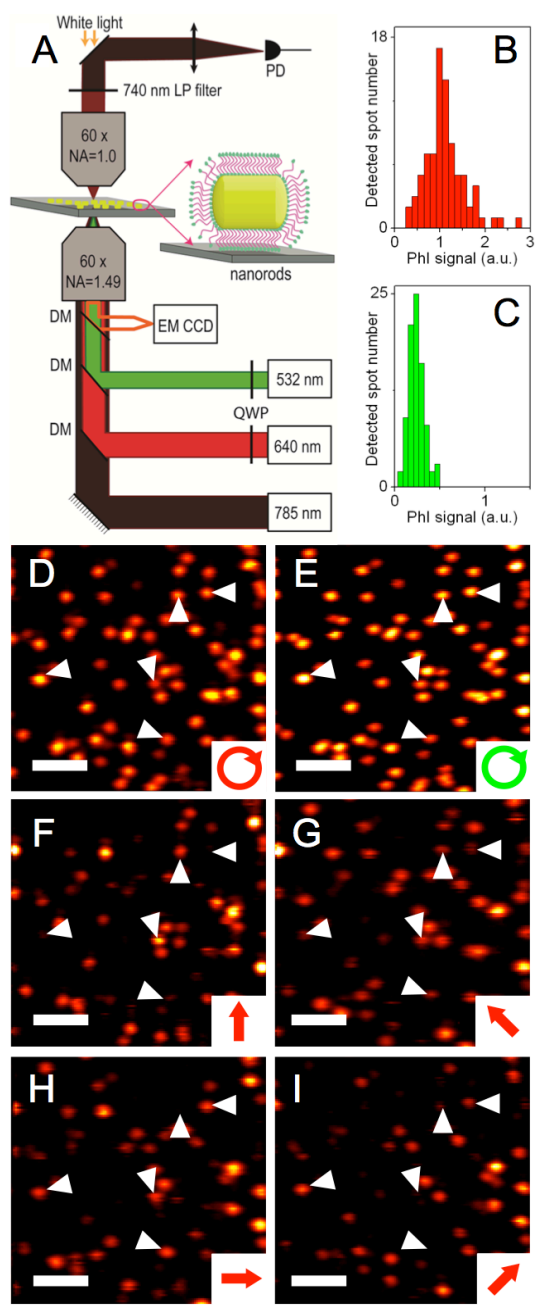

**Figure 3.** (A) Schematics of the two-color PhI microscope with excitations at 532 nm or 640 nm and a probe at 785 nm, (B and C) PhI images of nanorods (fraction with 642 nm peak absorption) excited with circularly polarized  (B) 532 nm and (C) 640 nm beams. (D and E): Corresponding PhI signal histograms.  (F-I) Same as (B°) but with linearly polarized 640nm excitation beams with orientations as indicated on the figures. Scale bars: 2μm. Arrowheads indicate the likely orientations of the in-plane projection of single nanorods long axis.

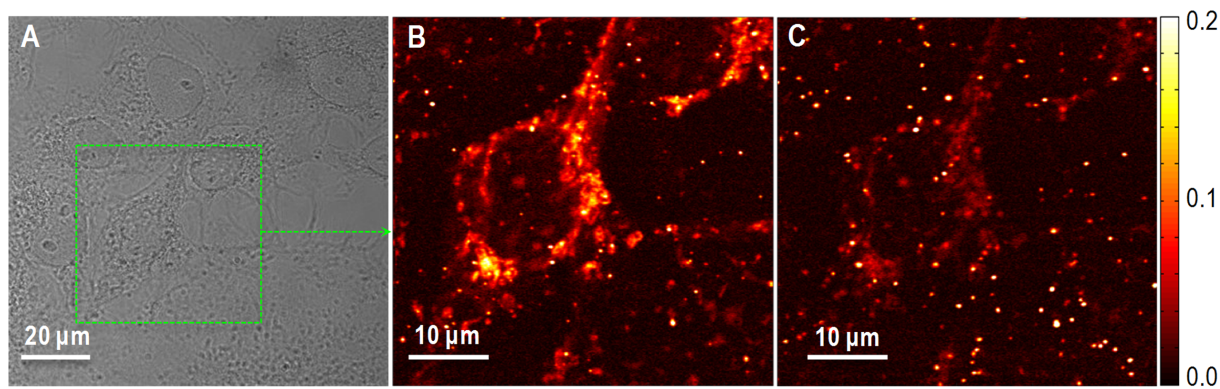

**Figure 4.** (A) White light and (B and C) PhI images of COS 7 cells incubated with nanorods under (B) 532 and (C) 640 nm excitation. PhI images recorded under red excitation show very weak mitochondrial background signals compared to that acquired under green excitation.

**The table of contents entry**

**We report the synthesis, sorting and characterization of mono-disperse gold nanorods with overall dimension around 10 nm in length and less than 6 nm in diameter.** They display strong and tunable red-shifted plasmon resonance, in a region where cellular absorption is reduced. A dual color photothermal microscope is developed to demonstrate that these nanorods are promising single molecule probes for cellular imaging in the near infrared.

**Keyword:** photothermal microscopy, gold nanorods, plasmonic materials, cell imaging, density gradient ultrahigh centrifugation

Edakkattuparambil Sidharth Shibu[1,2], Nadezda Varkentina[1,2], Laurent Cognet[1,2] and Brahim Lounis[1,2,*]

**Small gold nanorods with tunable absorption for photothermal microscopy in cells**

**ToC figure**

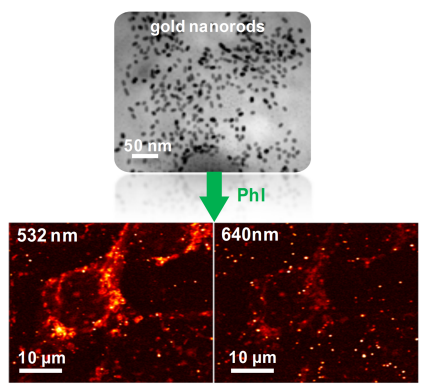